Graphical Abstract:

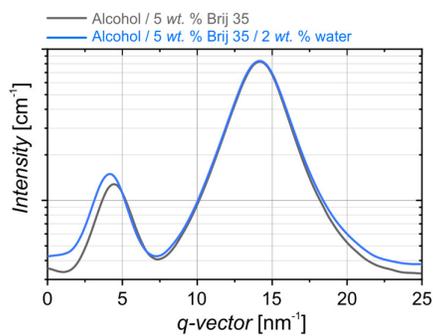

# Solvation of Nonionic Poly(Ethylene Oxide) Surfactant Brij 35 in Organic and Aqueous-Organic Solvents

*Jure Cerar*,[a] *Andrej Jamnik,*[a] *István Szilágyi*,[b] and *Matija Tomšič*.[a,*]

[a] Faculty of Chemistry and Chemical Technology, University of Ljubljana, Večna pot 113, SI-1000 Ljubljana, Slovenia.
[b] MTA-SZTE Lendület Biocolloids Research Group, Interdisciplinary Excellence Center, Department of Physical Chemistry and Materials Science, University of Szeged, H-6720 Szeged, Hungary

* Correspondence e-mail: Matija.Tomšič@fkkt.uni-lj.si; tel.: +386 1 479 8515 (M. Tomšič).

**Abstract**

*Hypothesis*: By combining the experimental small- and wide-angle x-ray scattering (SWAXS) method with molecular dynamics simulations and the theoretical 'complemented-system approach' it is possible to obtain detailed information about the intra- and inter-molecular structure and dynamics of the solvation and hydration of the surfactant in organic and mixed solvents, *e.g.*, of the nonionic surfactant Brij 35 ($C_{12}E_{23}$) in alcohols and aqueous alcohol-rich ternary systems. This first application of the complemented-system approach to the surfactant system will promote the use of this powerful methodology that is based on experimental and calculated SWAXS data in studies of colloidal systems. By applying high-performance computing systems, such an approach is readily available for studies in the colloidal domain.

*Experiments*: SWAXS experiments and MD simulations were performed for binary Brij 35/alcohol and ternary Brij 35/water/alcohol systems with ethanol, *n*-butanol and *n*-hexanol as the organic solvent component at 25 °C.

*Findings*: We confirmed the presence of solvated Brij 35 monomers in the studied organic media, revealed their preferential hydration and discussed their structural and dynamic features at the intra- and inter-molecular levels. Anisotropic effective surfactant molecular conformations were found. The influence of the hydrophobicity of the organic solvent on the hydration phenomena of surfactant molecules was explained.

## 1. Introduction

Surface-active agents or surfactants are amphiphilic compounds that show a strong tendency to self-assemble at the interfaces of immiscible phases and form supramolecular assemblies such as micelles, bilayers, etc. The nature of such assemblies strongly depends on the surfactant's molecular geometry and its surroundings. Deepening our understanding of the surfactant–solvent interactions and their effects on the structure of the system at the molecular level also improves our understanding of the system's structure–function relationship and aids the development of novel functional designs or novel application-oriented solvents or solvent mixtures. A combination of molecular dynamics (MD) simulations of structurally realistic model systems supplemented with the so-called 'complemented-system approach' for the x-ray scattering-intensity calculations [1, 2], which was developed in our lab, and the experimental small- and wide-angle x-ray scattering (SWAXS) technique is a powerful tool in this respect [2-7]. In the present study we successfully applied such a theoretically and experimentally grounded approach to study a nonionic surfactant system. This is the first time that the complemented-system approach has been applied to a system with relatively large molecules for treatment by computer simulation methods, *i.e.*, a surfactant system in the colloidal domain.

Among the various types of nonionic surfactants, the polyethylene oxide mono-ethers ($C_xE_y$; also known as PEO or POE surfactants) are some of most extensively used and studied [8-11]. In this study we used the polyoxyethylene (23) lauryl ether, which is better known under its commercial name Brij 35 or label $C_{12}E_{23}$, the latter indicating that the surfactant molecule has a relatively large hydrophilic head (twenty-three oxyethylene units) compared to its rather short alkyl-tail (eleven $CH_2$ units and one $CH_3$ unit). The Brij 35 molecule has a predominantly hydrophilic character – its HLB value is 16.9. While the structures of binary aqueous mixtures of Brij 35 have already been extensively studied [12-19], we could find only a few studies of Brij 35 in alcohol solvents [14, 20]. These latter systems are, for example, especially interesting in chromatography and phase-extraction applications and will be the main focus of this study. The phase diagrams of *n*-alcohol/Brij 35 systems [18, 21] reveal that in pure short-chain alcohols the solubility of Brij 35 is relatively high, whereas it decreases in the pure long-chain alcohols, but markedly increases in the presence of a small concentration of water (molar ratio of water/Brij 35 used in this study was approximately 26).

Our previous experimental study of Brij 35 in pure alcohols and in alcohol-rich aqueous systems indicated that the Brij 35 molecules in such systems are in the form of monomers [17]. However, that study could not provide detailed information about the solvation of Brij 35 monomers, but only allowed some circumstantial speculations about the effects of aqueous hydration of the surfactant molecules. Therefore, the main objective of the present work was to provide results on surfactant hydration and the structure and dynamics of the system at a molecular level of detail utilizing the complemented-system approach [1, 2]. This approach has already proven to be very useful in studies of pure organic solvents (alcohols, aldehydes, acids) [4, 7, 22, 23] and more recently also in the study of the structure-viscosity-dynamics relationship in a structurally and dynamically versatile aqueous system of *tert*-butanol across the whole composition range [2]. We now focus on the Brij 35 surfactant in amphiphilic organic solvents, *i.e.*, the short-chain *n*-alcohols with an increasing hydrophobic tail length: ethanol (EtOH), butan-1-ol (BuOH) and hexan-1-ol (HexOH). Although their molecules are very similar in terms of chemical structure, these media are different in their physico-chemical nature – by increasing the length of the alcohol tail, they represent a solvent that is increasingly hydrophobic in nature. In aqueous ternary Brij 35 systems where the micelles were observed, EtOH behaved as a structure-breaking agent, BuOH exhibited properties characteristic of cosurfactants, and HexOH and larger *n*-alcohols behaved like an oil penetrating the hydrophobic micellar core [18]. The present results revealed the changes in hydration of the surfactant molecules, the changes in the structure and the dynamics of the systems with respect to the size of the *n*-alcohol molecules. They can be generalized to similar systems of predominantly hydrophilic nonionic surfactants of the $C_nE_m$ type (and to some degree also to pure polyethylene glycol molecules) in such weakly amphiphilic organic solvents with considerable hydrophobic character, i. e., to the binary and ternary systems with similar phase-behaviour in the organic solvent-rich part of the phase diagram as was found for the studied systems (see Fig. 1 in ref. [18]).

## 2. Experimental and Methods

This study is based on the results of SWAXS measurements for binary systems containing alcohols (ethanol, *n*-butanol or *n*-hexanol) as solvents and 5 *wt.* % of nonionic surfactant Brij 35 ($C_{12}E_{23}$), and the ternary system samples also containing 2 *wt.* % of water at 25 °C. These concentrations of surfactant and water were selected because they are low enough to avoid macroscopic phase separation in these systems and at the same time this water concentration is high enough to provide reasonable statistics in simulated results. Similarly this specific surfactant and the systems were selected as we aimed to obtain the complementary more detailed information on solvation in these systems in respect to our previous experimental study [17]. The obtained results were used to check the accuracy of the molecular models by comparing them with the calculated SWAXS intensities from the molecular-dynamics simulation results and the theoretical complemented-system approach. Afterwards the simulated model-based results were used to interpret the structural, viscosity and dynamic properties of the studied systems at the intra-, inter- and supra-molecular levels of detail.

We give full details of the Materials in the following paragraph, but those of Small- and Wide-Angle X-ray Scattering Measurements [24, 25], Molecular Dynamics Simulations [26-32], Calculation of the Self-Diffusion Coefficients [33], Calculation of the Kirkwood-Buff

Integrals [34] and the Calculation of the X-Ray Scattering Intensity [1, 2, 33-42] based on the complemented-system approach [1, 2] in the online Supporting Information (SI).

The nonionic surfactant polyoxyethilene(23) lauryl ether ($C_{12}E_{23}$; Brij 35; Seppic, purity ≥ 99 %), ethanol (ECP, purity ≥ 99.8 %), butan-1-ol (Fluka Chemica, purity ≥ 99.5 %), and hexan-1-ol (Fluka Chemica, purity ≥ 98 %) were used as purchased without further purification. Demineralized water distilled in a quartz bi-distillation apparatus (Destamat Bi18E, Heraeus; specific conductance of the water was less than $6.0 \times 10^{-7}$ $\Omega^{-1}cm^{-1}$) was used to prepare the samples containing small concentrations of water. All samples were prepared directly by weighing. They were then tightly sealed, put into an ultrasonic bath and heated to ~ 50 °C for approximately 30 min to ensure that the Brij 35 was fully dissolved. Throughout the text the concentrations are presented as mass fractions, $w$.

## 3. Results and Discussion

We present and discuss the experimental and theoretical results for the binary Brij 35/alcohol and ternary Brij 35/alcohol/water systems with ethanol, *n*-butanol and *n*-hexanol as the alcohol component. The concentration of the Brij 35 surfactant in all samples is 5 *wt.* %, while in the ternary systems there is also 2 *wt.* % of water.

### 3.1. Experimental and Calculated SWAXS Data

The experimental and calculated SWAXS data are presented and compared on an absolute scale in Fig. 1. It is important to emphasize that for the confirmation of the validity of the theoretical models, which is checked through the agreement between the calculated scattering intensities and the experimental SWAXS data, the agreement in the position and the actual appearance of the individual scattering peaks is much more important than the agreement in the absolute values of the scattering intensities. The former is strongly dependent on the length scales and the geometry of the scattering objects, *i.e.*, the correct reproduction of the structural features or segments in the model system, whereas the latter depends on the correctness of the consideration of the scattering contrast, which is usually difficult to achieve in molecular models. The calculated scattering-intensity data presented in Fig. 1 reproduce all the features observed in the experimental data. Both the pronounced peaks at very low-*q* values (even the upturn in the scattering intensities) and their positions are reproduced very well. Correspondingly, we can conclude that the calculated scattering-intensity functions show very good agreement with the experimental data and confirm the validity of the models used in our study.

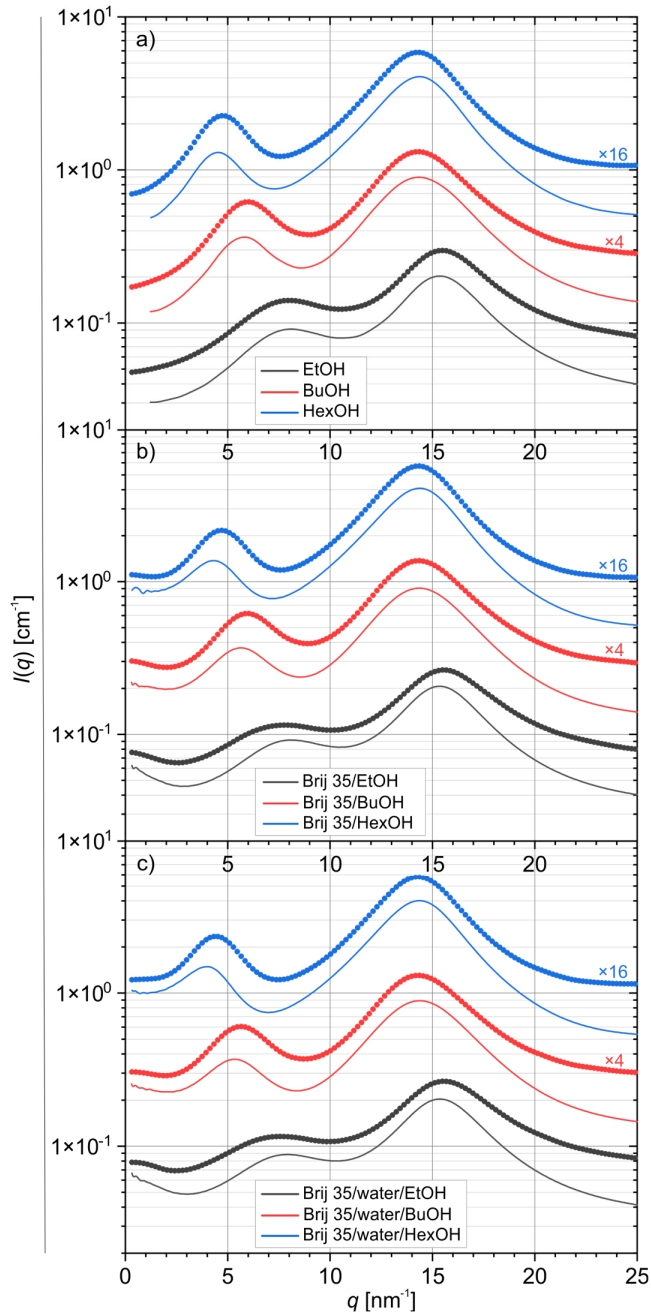

**Fig. 1.** The experimental (symbols) and calculated (lines) SWAXS intensities of (a) pure alcohols, (b) 5 *wt.* % Brij 35/alcohol binary systems, and (c) 5 *wt.* % Brij 35/2 *wt.* % water/alcohol ternary systems. Different alcohols are marked as: EtOH (black), BuOH (red), and HexOH (blue). For the sake of clarity, the data for BuOH and HexOH systems were shifted upwards, *i.e.*, multiplied by 4 and 16, respectively.

The results for the pure alcohol systems presented in Fig. 1a were already thoroughly discussed in several previous studies [6, 7, 43] and are presented here as a benchmark for our further discussion. To recapitulate, there are two typical scattering peaks observed in the SWAXS curves of the pure alcohols. The first is the so-called inner scattering peak that is positioned at the value of the scattering vector around 5 nm$^{-1}$. As this peak primarily arises from the correlations between the neighbouring 'chains' of sequentially hydrogen-bonded hydroxyl groups (the so-called –OH skeletons), which form the hydrophilic regions in the system that are separated by the hydrophobic regions, its position actually depends on the alcohol's hydrocarbon tail length. The second is the outer scattering peak that is positioned at the value of the scattering vector around 15 nm$^{-1}$ and arises from the inter- and intra-molecular correlations within the hydrophobic regions (alcohol hydrocarbon tails). When 5 *wt.* % of the surfactant Brij 35 is present in the alcohol the positions and the absolute scattering intensities of the alcohol's inner and outer peaks in the SWAXS curves remain practically unchanged, as can be seen in Fig. 1b. This indicates that the bulk alcohol structure remains practically unaffected by the presence of the surfactant. However, we can clearly see higher absolute intensity values and even an upturn in the scattering curves in the very low-$q$ region (below ~2 nm$^{-1}$) in comparison with the results for the pure alcohols, which reflects the presence of larger scattering structures in these binary systems. The experimental data also reveal that this upturn does not seem to change much with an increase in the hydrophobic nature of the solvent (from ethanol to hexanol).

Interestingly, the presence of a small concentration of water (2 *wt.* %) leaves the outer scattering peak and the very low-$q$ upturn practically unchanged, whereas the position of the inner scattering peak Fig. 1c shifts slightly towards the lower-$q$ values. It seems that such a shift is a little more pronounced in the case of BuOH and HexOH. At this stage we can only state that this is clearly an effect of the presence of water in such ternary systems, although the details of its influence on the structure remain hidden. The model-based results that are presented and discussed later in the text will help to reveal it.

### 3.2. Inter-Molecular Structure

According to our previous experimental small-angle x-ray scattering (SAXS) study of these systems based on the Indirect Fourier Transformation of the SAXS data [17], we were expecting to find the presence of solvated and hydrated monomers of the nonionic Brij 35 surfactant. However, as we applied the molecular models in the present study, our aim was to obtain detailed information about the effective molecular configuration of the nonionic surfactant and its solvation and hydration. Such information, which is crucial for an understanding of the intermolecular interactions, is not obtainable just from the experimental studies.

The structural situation observed in the modelled Brij 35/HexOH and Brij 35/water/HexOH system is depicted in Fig. 2. A similar finding was made for the systems with EtOH and BuOH as the organic solvent in Figure S1 in the online Supporting Information (SI). This confirms the presence of solvated and hydrated Brij 35 monomers in these systems – the individual surfactant molecules are surrounded by at least one 'layer' of alcohol molecules, which is also obvious from the radial distribution functions discussed later in the text (see Fig. 4d). From the results on the ternary system depicted in Fig. 2b, we can conclude that a large fraction of the water molecules are hydrating the Brij 35 hydrophilic heads, whereas the rest incorporate themselves with the hydrogen bonds within the alcohol –OH group's skeletal structure. A statistical analysis of the results revealed that the fraction of the water molecules hydrating the Brij 35 molecules is $47 \pm 1$) %, ($55 \pm 1$) %, and ($57 \pm 1$) % for the EtOH, BuOH, and HexOH systems, respectively.

In order to extract additional structural information from the calculated SWAXS intensities, we turn our attention to some of the partial scattering contributions presented in Fig. 3 (see also Figure S2 in the online SI). The details of how we obtained these contributions according to the complemented-system approach are described elsewhere [1, 7]. For the present discussion, it suffices to know that we can always represent the total calculated scattering intensity of the modelled system as the sum of a number of the partial scattering contributions, where some of them represent the direct partial contributions of the atoms/pseudo-atoms of an individual type (or a group of atoms/pseudo-atoms that, for example, represent an individual molecular type) and the others the cross-term contributions (interference terms) of the atoms/pseudo-atoms of two (or more) different types. In such a way, we can calculate the contributions of only a certain type or types of molecules, which provide similar information to the contrast-matching experiments typical for small-angle neutron scattering [44]. Therefore,

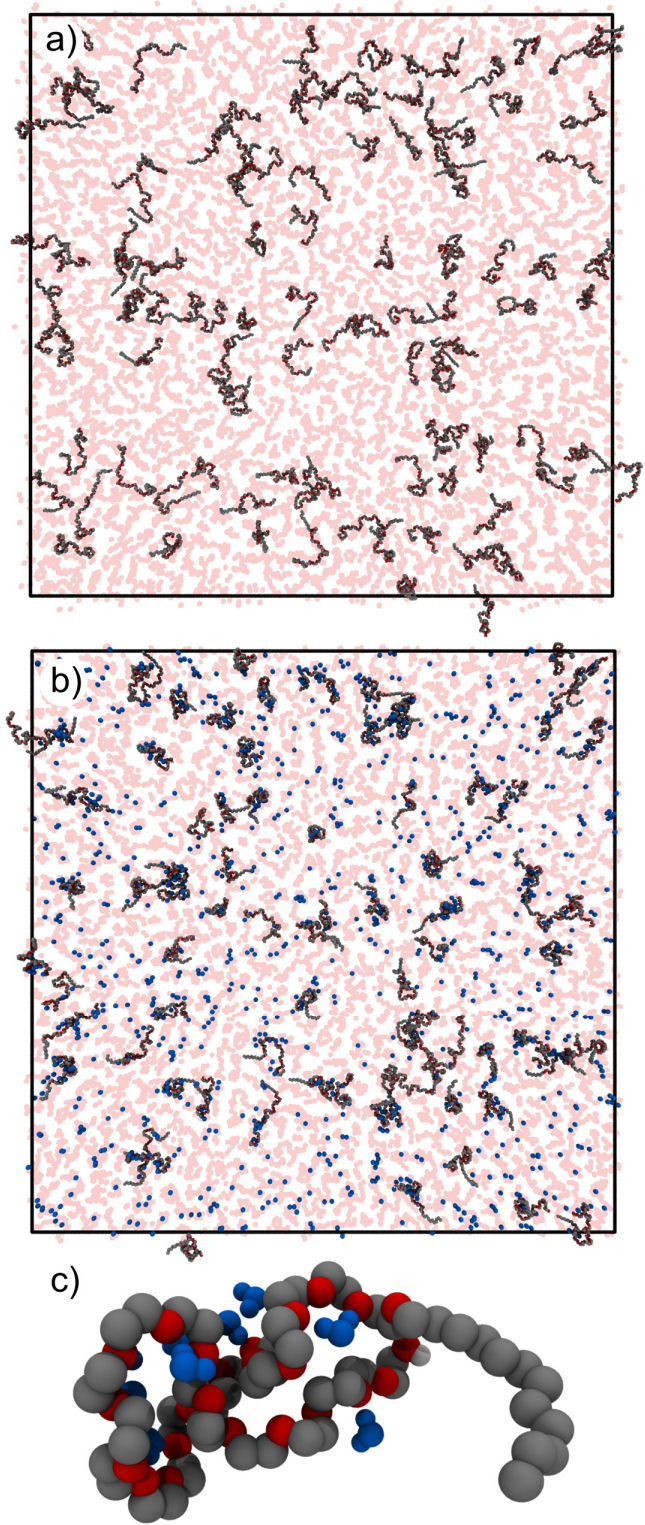

**Fig. 2.** The visualization of 1 nm thick 40 nm × 40 nm slice of the MD-simulation box for the (a) 5 wt. % Brij 35/HexOH, and (b) 5 wt. % Brij 35/2 wt. % water/HexOH systems, where the alcohol molecules are omitted for the sake of clarity – the pale red areas indicate the hydrophilic regions formed by the sequentially hydrogen-bonded alcohol hydroxyl groups; the water

molecules are depicted in blue. (c) A typical single Brij 35 molecule with its hydrated water is shown.

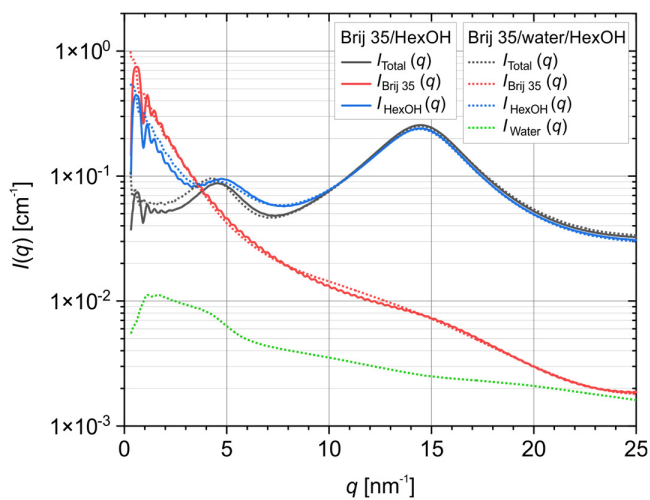

**Fig. 3.** The partial contributions (not smeared) to the theoretical SWAXS intensities of: Brij 35 (red), HexOH (blue), and water (green) molecules, as well as the total scattering intensities (black) for the Brij 35/HexOH (full lines) and Brij 35/water/HexOH (dotted lines) systems. The scattering cross terms are not shown.

we usually represent this part of our approach as the 'theoretical analogy to the contrast-matching experiments', which are otherwise not possible in the experimental SAXS studies.

Some interesting features can be observed in the data in Fig. 3. The partial contribution of Brij 35 to the total scattering shows a sharp upturn in the low-$q$ region, which indicates the presence of some large structures, a 'shoulder' at around 15 nm$^{-1}$, where typically the correlation lengths characteristic for the hydrophobic hydrocarbon tail regions are expressed (the outer scattering peak of the alcohols), and the considerable decrease of the scattering intensity with increasing value of the scattering vector. A similar low-$q$ scattering-intensity upturn can be seen in the scattering contribution of the alcohol, which can be explained as a consequence of the so-called Babinet's principle, *i.e.*, the contribution of the voids in the structure of the alcohol, which represent the space occupied by the surfactant molecules in the system. More interestingly, we can observe that the inner and outer peak positions arising from the alcohols (full and dotted blue curves) are practically unaffected by the presence of water in the system – this indicates that the water molecules in small concentrations do not change the bulk structure of the alcohol significantly. An important conclusion that follows from this observation and could not be made solely on the basis of the experimental data is the fact that the shift of the inner scattering peak (full and dotted black lines), which was also observed in the experimental and the calculated SWAXS data in Fig. 1c, is a consequence of the hydration of the Brij 35 hydrophilic heads and not a consequence of an eventual change in the structure of the organic solvent (bulk alcohol). This can be confirmed by a very interesting feature of the water's contribution to the total scattering in Fig. 3. This contribution shows a very broad peak in the range of the scattering-vector values from ~1 to ~5 nm$^{-1}$, which is rather untypical for water, as it is not observed in the scattering of the bulk water (*e.g.*, see the lowest scattering curve in Figure 1a in ref. [2]). This means that in this system the water molecules are correlated over rather long distances, which can only be explained as a consequence of the hydration of large and relatively stiff Brij 35 hydrophilic heads, *i.e.,* the hydrated water is to some extent 'arrested' within the hydrophilic region of the Brij 35 molecules. This also proves that a considerable fraction of the water molecules is indeed hydrating the Brij 35 molecules.

Further information about the structure and hydration can be obtained from the radial pair-distribution functions, $g(r)$ shown in Fig. 4. They were calculated with respect to the centre of mass (COM) of the modelled Brij 35 molecules. Therefore, in Fig. 4 there are four distinct $g(r)$ functions, *i.e.*, COM–water, COM–COM, COM–AlcOH and COM–AlcCH$_x$, where the latter two represent the radial arrangement of the alcohol hydroxyl heads (AlcOH) and the alcohol alkyl tails (AlcCH$_x$) around the COM of surfactant molecules, respectively. To facilitate the discussion, a schematic representation of molecular arrangement around the COM structure is presented in the upper part of Fig. 4, with the dotted concentric circles around the COM in the scheme corresponding to some characteristic correlation-length features of the individual $g(r)$ function.

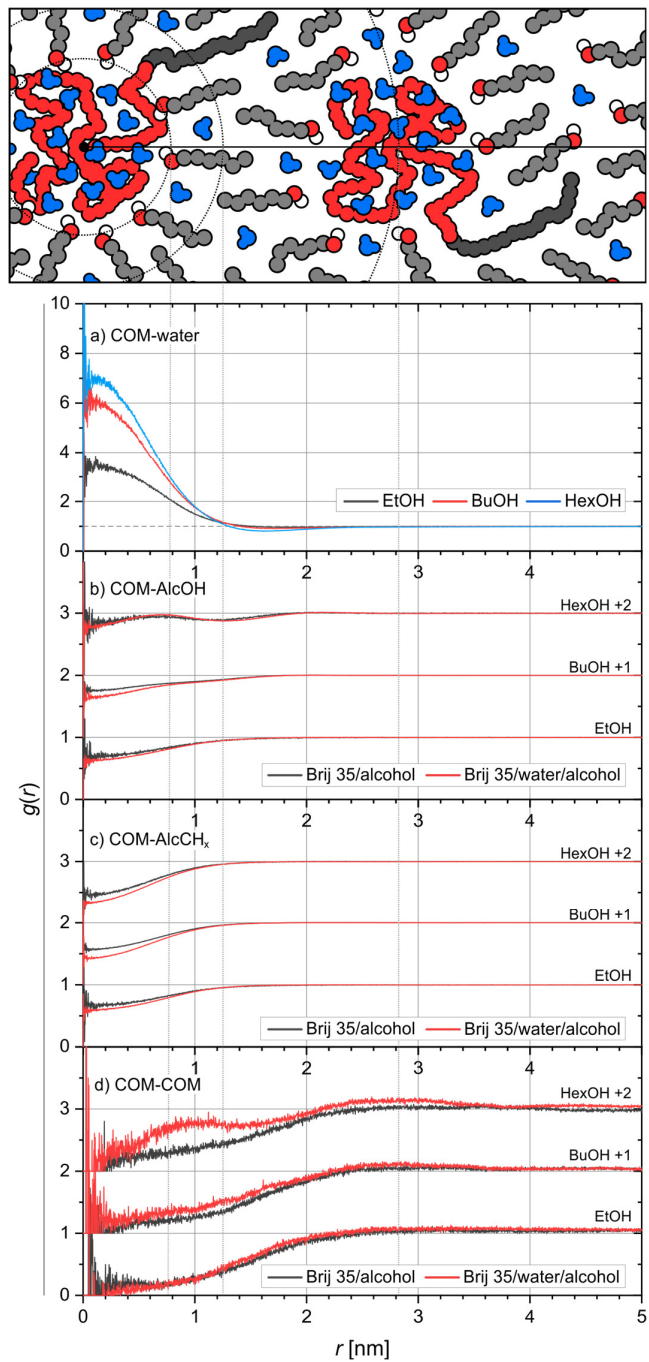

**Fig. 4.** *Top*: The schematic representation of molecular arrangements around the Brij35's centre-of-mass (COM). *Bottom*: The radial pair-distribution functions, $g(r)$, for (a) COM–water, (b) COM–AlcOH, where AlcOH corresponds to the –OH group of the alcohol molecule, (c) COM–AlcCH$_x$, where AlcCH$_x$ corresponds to the alkyl tail of the alcohol molecule, and (d) COM–COM. In (b), (c) and (d) the functions for Brij 35/alcohol system are plotted as black full lines and those for Brij 35/water/alcohol system as red full lines; the functions for EtOH are plotted on the original scale, but the ones for BuOH and HexOH are shifted upwards by the value of 1 and 2, respectively.

Possibly the most interesting structural feature connected to the hydration of the Brij35 hydrophilic heads can be clearly observed from the radial COM–water distribution in Fig. 4. It is obvious that the water molecules tend to preferentially arrange around the Brij 35's COM, which is in any case located very close to the COM of the Brij 35 hydrophilic head. The central peak in the $g(r)$ function in Fig. 4a grows with the increasing size of the alcohol alkyl tail and reveals that the hydration of the Brij 35's molecules is increasing with the increasing hydrophobic nature of the organic solvent. Accordingly, we can also observe in the COM–AlcOH and COM–AlcCH$_x$ distribution functions presented in Fig. 4b and 4c that these functions tend to show somewhat lower values at very short distances for the systems containing water. This feature is, of course, another consequence of an increase in the hydrophilicity of the surfactant hydrophilic-head regions due to their hydration, causing an even lower probability of finding the organic solvent molecules close to the COM, compared to the situation in the systems containing no water. More detailed discussion can be found in the online SI. The corresponding Kirkwood-Buff integral values for all $g(r)$ functions shown in Fig. 4 are also given in Table S1 in the online SI.

The COM–COM $g(r)$ functions presented in Fig. 4d provide an insight into the intermolecular correlations between the neighbouring surfactant molecules. They reveal that the surfactant molecules prefer to be solvated as monomers, even though in HexOH a peak at around 1 nm starts to appear, indicating a somewhat increased probability of closely neighbouring Brij 35 molecules; as these values of the $g(r)$ function are still below one in this range of distances the probability for such an organization of surfactant molecules is rather low. The most probable inter-molecular distances between the two neighbouring Brij 35 molecules are around 2.7 nm, as indicated by the highest value of the $g(r)$ function. Interestingly, this appears to be roughly the same for all the studied systems, regardless of the solvent. The addition of water to the system seems to 'stabilize' the surfactant monomers in the case of EtOH, whereas in parallel it seems to lead to a higher probability of finding neighbouring surfactant molecules at shorter distances in BuOH and HexOH. This could be explained as a consequence of the increasing hydrophilic nature of the hydrated surfactant heads in the otherwise more hydrophobic nature of the organic solvent. This is to be expected as these ternary systems are close to the phase-separation curves in the phase diagrams [18, 21].

### 3.3. Intra-Molecular Structure

The results on effective molecular conformations of the modelled Brij 35 molecules are depicted in Fig. 2. In general, the hydrophilic surfactant head tends to take up more-or-less coiled, compacted shapes, whereas the hydrophobic surfactant tail seems to protrude away from them into the organic solvent, as depicted schematically in the upper part of Fig. 5. Considering all the surfactant molecules in the simulation box, the intra-molecular spatial-distribution functions were obtained and are shown with the distribution *x-y* and *y-z* cross-section profiles in Fig. 5. We also calculated the spatial distribution of the water around the COM, which clearly shows intensive hydration of the hydrophilic surfactant head (blue curve).

It is clear from the intra-molecular spatial-distribution functions presented in Fig. 5 that the effective conformation of the Brij 35 molecule in these organic solvents resembles a lachrymiform shape, *i.e.*, the anisotropic drop-like shape (asymmetric *x-y* cross-section and symmetric *y-z* cross-section). Such a shape is, of course, a consequence of the dual nature of the surfactant molecule and the predominantly hydrophobic environment in the systems. The shape-anisotropy ratio, which represents the ratio between the x semi-axis and the y semi-axis, for this shape is 2.2, with a standard deviation of 0.5. Interestingly, this value is in perfect agreement with the value of 2.2 ± 0.7 that was reported for the random-walk polymer [45, 46]. Nevertheless, we have to clarify that in these systems the surfactant molecules show no preferential orientation and that these binary and ternary systems are actually optically isotropic in their nature.

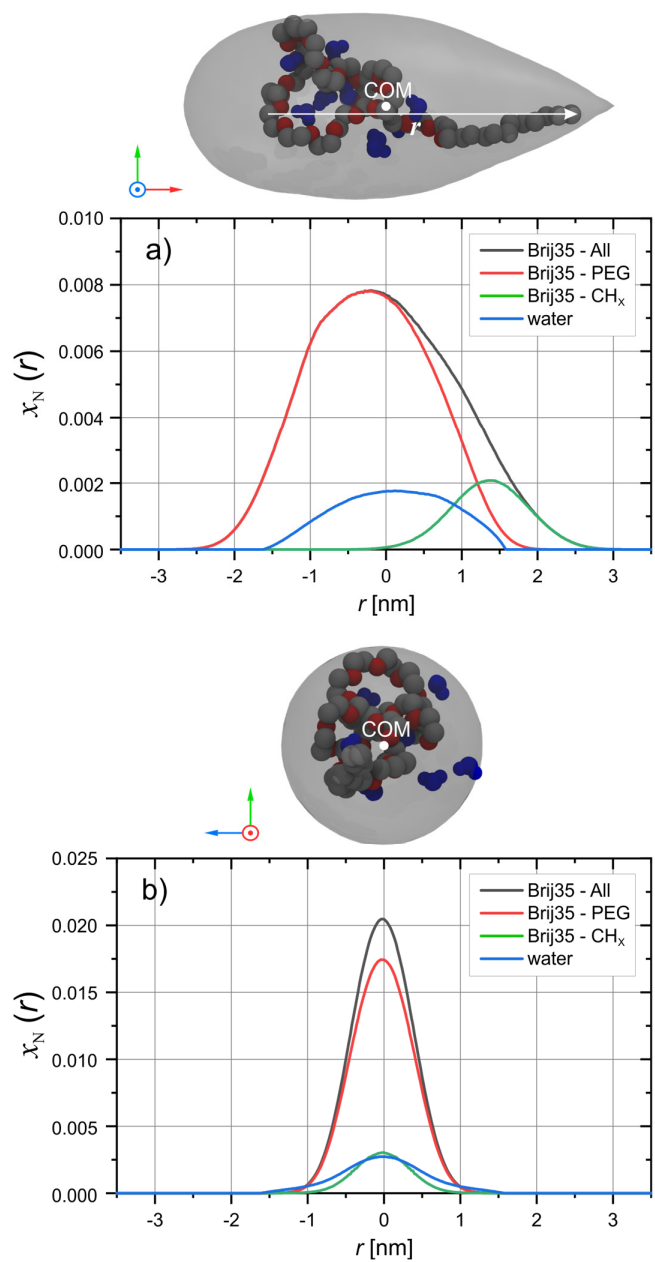

**Fig. 5.** *Top:* Schematic representation of the hydrated Brij 35 molecule and the two cross-sections views of the calculated intra-molecular spatial distribution function of Brij 35 for Brij 35/water/HexOH system. *Bottom:* The corresponding (a) *x-y* cross-section and (b) *y-z*-cross-section distribution profiles of solely the surfactant hydrophilic-head chain (red line), solely surfactant hydrophobic alkyl-tail chain (green line) and the whole surfactant molecule (black line) in the ternary system. The probability distribution of water around the COM is also presented (blue dotted line). The quantity $x_N$ corresponds to the number fraction of the individual molecules (molecular parts) or the probability of finding them at a specific distance $r$ in respect to the COM.

The overall effective shape of the surfactant molecules does not seem to depend substantially on the hydrophobicity of the organic solvent used in this study, but we can observe some changes in its dimensions in Fig. 6a. With an increase in the hydrophobic nature of the solvent the intra-molecular distribution $x$-$y$ cross-section profile seems to become slightly narrower, indicating a more compact shape of the surfactant hydrophilic head, whereas the $y$-$z$ cross-section distribution profile does not change significantly. This means that with an increase in the hydrophobicity of the solvent the hydrophilic part of the Brij 35 molecules shrinks slightly and becomes more compact to minimize the contact with the solvent. Such a compaction of the surfactant's effective molecular conformation can also be observed even more easily in the related distributions of the molecular end-to-end distances of the modelled surfactant molecules, which are presented in Fig. 6b. They clearly show that with an increasing hydrophobicity of the organic solvent (from EtOH to HexOH) the peak position of the distribution shifts towards shorter end-to-end distances corresponding to more compact molecular conformations. The Brij 35 molecule has a huge hydrophilic head in relation to its short hydrophobic tail and correspondingly a high HLB value (16.9) [47], reflecting its relatively strong hydrophilic character. Therefore, such behaviour is expected in an increasingly hydrophobic environment. It is discussed into more details in the online SI.

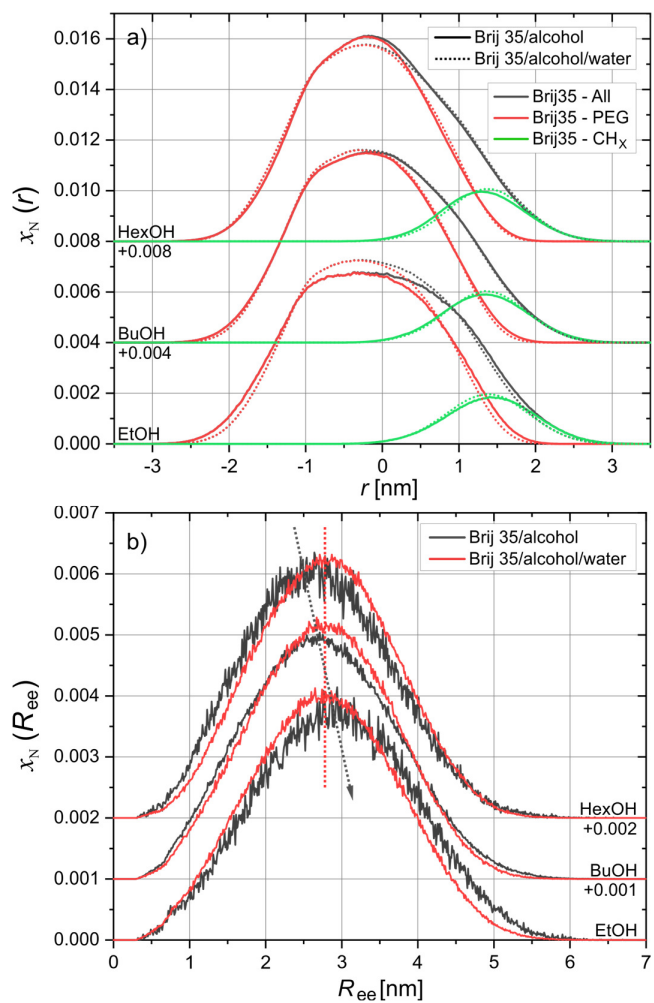

**Fig. 6.** The x-y cross-section intra-molecular distribution profiles of the surfactant molecules in the alcohol systems with (dotted lines) and without water (full lines): solely hydrophilic-head chain (red), solely hydrophobic alkyl-tail chain (green) and the whole surfactant molecule (black line). (b) The distribution function of end-to-end distance of the modelled Brij 35 surfactant molecule in the EtOH, BuOH and HexOH systems with and without water.

Moreover, when the surfactant's hydrophilic heads are hydrated they tend to adopt more globular shapes, which is a feature that was speculated on already in our previous experimental study of these systems [17], but is confirmed by the model-based results of the present study. This can be inferred from the more symmetric course of the distribution *x-y* cross-section profiles in Fig. 5c for the ternary systems containing a small concentration of water. Similarly, the presence of water in the studied systems causes the constant position of the peaks in the distribution of the end-to-end distances in Fig. 6b, irrespective of the kind of organic solvent. This can only be explained by an increasing fraction of water that hydrates the hydrophilic part of the surfactant molecules as the solvent is more and more hydrophobic, consequently making this part of the molecules bulkier in comparison with those in the systems without water. This then coincidentally cancels out the change in the position of the peak. The fact that in the EtOH system with water the peak in the distribution of the end-to-end distances is interestingly slightly shifted to shorter distances can be explained as a consequence of the intra-molecular bridging with the hydrating water molecules.

### 3.4. Structure-Dynamics Relationship

Finally, we pay attention to the dynamic properties of the system and relate them to the observed structural features. Such properties are particularly interesting in the context of the use of such systems in chromatography and similar applications. We calculated the self-diffusion coefficients of the molecules in the binary and ternary model systems based on the mean-square displacement using the Einstein relation [33]. The calculated coefficients for the alcohol, Brij 35 and water molecules are presented in Table 1 and reveal their translational mobility in the studied systems.

To check and validate our results, we first calculated the self-diffusion coefficients for pure alcohols (EtOH, BuOH, HexOH) and water. The value obtained with the TIP4P/2005 water model is comparable to the nominal value of the model of $2.1 \times 10^{-5}$ cm$^2$/s from the literature [38, 39, 48] and the reported experimental value of $2.30 \times 10^{-5}$ cm$^2$/s [39, 40]. For the pure alcohols modelled with the TraPPE-UA force field, the calculated self-diffusion coefficients are obviously somewhat overestimated compared to the reported experimental values, but they follow the correct trend: EtOH 0.98×10$^{-5}$ [49] and 1.01×10$^{-5}$ cm$^2$/s [50], BuOH 0.416×10$^{-5}$ [41] and 0.426×10$^{-5}$ cm$^2$/s [51], and HexOH 0.218×10$^{-5}$ cm$^2$/s [41]. However, the value for EtOH is comparable to that obtained from a more detailed OPLS model for EtOH, i.e., 1.406×10$^{-5}$ cm$^2$/s [42].

**Table 1.** Diffusion coefficients, $D$, of alcohol, Brij 35, and water in different systems.

| System | $D$(Alc) [$10^{-5}$ cm$^2$/s] | $D$(Brij 35) [$10^{-5}$ cm$^2$/s] | $D$(water) [$10^{-5}$ cm$^2$/s] |
|---|---|---|---|
| water | | | 2.283 ± 0.042 |
| EtOH | 1.385 ± 0.007 | | |
| EtOH/water | 1.282 ± 0.025 | | 1.208 ± 0.081 |
| EtOH/Brij 35 | 1.256 ± 0.028 | 0.272 ± 0.085 | |
| EtOH/Brij 35/water | 1.200 ± 0.013 | 0.269 ± 0.052 | 0.97 ± 0.042 |
| BuOH | 0.607 ± 0.015 | | |
| BuOH/water | 0.615 ± 0.012 | | 0.632 ± 0.008 |
| BuOH/Brij 35 | 0.550 ± 0.018 | 0.175 ± 0.059 | |
| BuOH/Brij 35/water | 0.554 ± 0.018 | 0.179 ± 0.052 | 0.515 ± 0.023 |
| HexOH | 0.348 ± 0.020 | | |
| HexOH/water | 0.341 ± 0.018 | | 0.349 ± 0.002 |
| HexOH/Brij 35 | 0.309 ± 0.022 | 0.136 ± 0.029 | |
| HexOH/Brij 35/water | 0.299 ± 0.020 | 0.106 ± 0.045 | 0.289 ± 0.007 |

We were also able to find some data on the diffusion coefficient for the binary system EtOH/water, e. g. the experimental value of the (tracer) diffusion coefficient of EtOH at 98.3 wt. % EtOH of 1.022×10$^{-5}$ cm$^2$/s [52] and the estimate of the experimental NMR diffusion coefficient value of water from the reported diagram, which is about 0.9×10$^{-5}$ cm$^2$/s [48]. The former value was not determined at the concentration of EtOH we studied, but the overestimation in our calculated value is obvious. However, it seems to agree well with the calculated values obtained by Klinov and Anashkin using different models [53]. At this point, we must also mention an interesting study by Su *et al* [54] on the diffusion of water in *n*-alcohols measured by a novel single microdroplet method, which gave the following values for the diffusion coefficient of water in EtOH, BuOH and HexOH: 1.22×10$^{-5}$ cm$^2$/s [55], 0.56×10$^{-5}$

cm$^2$/s [55] and 0.35×10$^{-5}$ cm$^2$/s [54], respectively. Interestingly, these values are quite close to the calculated values for water obtained in the corresponding binary systems in our study.

Unfortunately, we could not find any experimental or calculated literature data on the self-diffusion coefficients of the surfactant Brij 35 and other studied molecules directly in the ternary systems under investigation. Nevertheless, we can mention some interesting studies on the thermal diffusion of poly(ethylene oxide) in ethanol/water mixtures [56-58] and on the self-diffusion of polyethylene glycol in water [59]. Although these are diffusion results on somewhat different systems from those studied here, it is interesting to see that the experimentally determined values of the self-diffusion coefficient of polyethylene glycol with the molecular mass of 943.1 g mol$^{-1}$ (21 PEO units) and 1603.9 g mol$^{-1}$ (36 PEO units) in water at infinite dilution are 0.334 ×10$^{-5}$ cm$^2$/s and 0.267 ×10$^{-5}$ cm$^2$/s, respectively (Brij 35 has 23 PEO units) [59]. These values were obtained in water as solvent and are slightly higher, but still in the same order of magnitude as those we obtained in the studied alcohols.

To conclude this brief literature review, although in our study we find some overestimation of the calculated self-diffusion coefficients in an absolute quantitative sense, they can certainly still be used qualitatively to explain some dynamical trends and the structure-dynamics relationship in the systems studied. Therefore, the discussion of these results deserves some more attention and is given below. The presence of Brij 35 surfactant molecules in the alcohol causes a lowering of the mobility of the alcohol molecules. This can be explained by the fact that the surfactant molecules hinder the alcohol molecules traversing the system. In alcohols, large amphiphilic parts of the Brij 35 molecules are strongly solvated, thus giving rise to hydrogen bonding with the alcohol molecules finding themselves in the vicinity of the solvation shells. Interestingly, the presence of water in a small concentration (2 *wt.* %) does not seem to influence the mobility of the alcohol molecules significantly, with the exception of the ethanol, which is the most hydrophilic alcohol in the studied series. However, the mobility of the water is significantly reduced, as a consequence of its hydrogen bonding to the –OH skeletal structure of the alcohol molecules. Consequently, their mobility is limited by the slower reordering of the –OH skeletal structure and by the relatively large hydrophobic regions between the –OH skeletons that represent an additional hindrance to the mobility of the hydrophilic molecules. We have described such influences into more detail in our recent study of the alcohol *tert*-butanol and water binary systems [2]. With an increasing size of the alcohol molecule (alkyl tail) the hydrophobic regions become larger and the reordering of the –OH skeletons becomes slower, which in turn further decreases the mobility of the water, as reflected in the even lower values of the self-diffusion coefficients given in Table 1.

Similarly, the mobility of the Brij 35 molecules is also lowered with an increase in the alcohol's alkyl tail length, as in this direction the viscosity of the solvent increases. Interestingly, the presence of water in the ternary systems does not cause any noticeable change in the mobility of the Brij 35 molecules (with the exception of the EtOH system), even though we would expect some decrease in the mobility because of the hydration of the surfactant molecules. This might be connected with the changes in the size of the surfactant molecules due to the interplay of the effects of the hydrophobic environment and the extensive hydration. The structural and conformational analysis showed that the water molecules form hydrogen bonds with the Brij 35 hydrophilic heads. Therefore, we would expect a correlated mobility of the two species, so lowering the mobility of both. On the other hand, the mobility of water molecules in ternary systems with Brij 35 and alcohols is actually much lower than that found in the binary alcohol/water systems. Two 'kinds' of water molecules are actually present in the ternary systems – the water in the 'bulk' alcohol environment and that in the hydration shell around the surfactant molecules. The mobility of the latter is to a great extent determined by the mobility of the surfactant and is for sure much lower than that of the bulk water.

## 4. Conclusions

The results presented in this work demonstrate for the first time that it is possible to successfully combine the experimental SWAXS and the MD computer-simulation methods with scattering-intensity calculations using the complemented-system approach [1, 2] to provide a detailed insight into the structure and dynamics of a modelled binary and ternary nonionic surfactant system. In this way we promote this approach as a successful novel method in the field of SWAXS data interpretation [7, 36, 60-71] that can also reveal details of the structure and dynamics for colloidal systems. Without an insight into the model-based simulation results, such details at the intra- and inter-molecular levels could not be obtained. The methodology also exploits the theoretical analogy of the so-called 'contrast matching', which is well known in experimental small-angle, neutron-scattering studies, in the x-ray scattering domain, where otherwise it cannot be applied using classic approaches [71]. This study also demonstrates that the TraPPE-UA force-field model [72, 73] can be successfully 'transferred' to describe the properties of larger molecules like nonionic surfactants, even though originally this model was neither parametrized nor optimized for such systems.

The results confirmed the presence of Brij 35 surfactant monomers in pure alcohol solutions. They revealed an interesting effect of increasing the hydrophobic nature of the organic solvent, as well as the presence of water in the system, on the surfactant's molecular conformations. With the calculated partial SWAXS contributions, we were able extract the origins of the scattering features in the experimental SWAXS curves and to explain the link between the changes in the scattering intensities and the structure of the system. The radial pair distribution functions clearly indicated intermolecular solvation and hydration phenomena. The main conclusions were also supported by information about the intramolecular structure of the surfactant molecules obtained from the spatial 3D distribution functions and molecular end-to-end distances. The dynamic properties of the system were discussed in terms of the molecular mobility inferred from the calculated self-diffusion coefficient values.

We can conclude that this research importantly complements our previous, purely experimental studies of these and related binary and ternary systems [17, 18], which could not offer much more for the alcohol-rich compositions than to suggest the presence of Brij 35 molecules in the form of monomers in such systems. Its main novel contribution is the detailed insight into the structural and dynamic properties of the surfactant molecules connected to their solvation and hydration in the predominantly hydrophobic environment at the intra-, inter- and supra-molecular levels. The effective conformation of the Brij 35 surfactant molecules in these systems resembles a typical drop-like shape. Because of its relatively large hydrophilic PEO moiety compared to its rather short alkyl tail, which is readily solvated in a hydrophobic organic alcohol, the PEO portion of the molecule represents a locally segregated hydrophilic domain within the predominantly hydrophobic environment, as shown schematically in Fig. 4. When water is also present in the system, this hydrophilic domain becomes markedly hydrated, a feature that is enhanced as the hydrophobic character of the alcohol increases.

Even though there are many experimental [12-15, 17, 18] and theoretical [16, 19] studies of the structure of the nonionic surfactant Brij 35 in water available in the literature, we have not found a similar study for Brij 35 in alcohol solvents that would deal with the monomer surfactant molecules and their interactions with the solvents in the details presented here. The solvation and hydration of macromolecules and molecular assemblies (e.g., various molecular probes [74], proteins [75, 76], DNA [77, 78], micelles [79, 80], microemulsion systems [81, 82]) is a very attractive and important research topic. Some of the systems investigated in our study could arguably be considered as somewhat less interesting as they lack any larger supra-molecular assemblies (*e.g.,* micelles or similar), but they are still very interesting from the viewpoint of applications in chromatography [83-85], reaction media [86], polymer melts, delamination processes [87, 88], etc. Furthermore, we believe that the presented results can be generalized to similar systems of nonionic surfactants of type $C_xE_y$ with predominant hydrophilic character, as well as to poly(ethylene glycol) block co-polymer systems with predominant PEO unit, and to some extent even to some pure poly(ethylene glycol) polymer systems. The common feature of such systems would be PEO units of similar size, or better similar predominant hydrophilic character, solvated in a weakly amphiphilic organic environment with significant hydrophobic nature. These units, would represent local hydrophilic "pools" in the hydrophobic environment, which would be significantly hydrated if some water were also present.

Such educated predictions would be even more justified for the systems that show similar composition dependent behavior in the organic-solvent-rich part of the phase diagrams as the systems studied (see Fig. 1 in ref. [18]). As follows from the course of the single-phase region in these phase diagrams and from our detailed results on the solvation phenomena in these systems, the presence of some water actually induces the increased solubility of the hydrophilic PEO units in such hydrophobic organic solvents. The explanation for this behavior most likely lies in the extensive hydration of the hydrophilic PEO units, which affects their flexibility and dynamics and contributes to the stability of the single-phase state of the system. With respect to the surfactant concentration in the system, we are confident that the presented results on the structure and dynamics of solvation and hydration are indicative also for the systems at lower surfactant concentrations – the surfactant can be expected to be in a form of monomers there as well. With increasing concentration of the surfactant (and the concomitant increase in the concentration of water in the system to remain in the single-phase region of the phase-diagram), one would expect a gradual onset of molecular aggregation of such PEO units and the corresponding increase in the size of the locally hydrophilic PEO domains in such organic solvent-rich systems. However, these are only predictions, as the concentration and temperature dependence of the situation in such systems is beyond the scope of this study. Such dependencies could of course be interesting for future studies.

With the rapid development and increase of computational power over recent years, our goal is to apply our methodology to even larger and more complex molecular systems, *e.g.,* ionic-liquids, supra-molecular self-assembly systems, polymer melts, and others.

**DECLARATION OF COMPETING INTEREST**

There are no conflicts to declare.

**ACKNOWLEDGEMENTS**

We acknowledge financial support from: the Slovenian Research Agency (research core funding No. P1-0201, the project No. N1-0139 'Delamination of Layered Materials and Structure-Dynamics Relationship in Green Solvents') and the Hungarian National Research, Development and Innovation Office (projects no. SNN131558). We are most grateful to Prof. Otto Glatter for his generous contribution to the instrumentation of our Light Scattering Methods Laboratory in Ljubljana.

**CRediT authorship contribution statement**

**Jure Cerar:** Data curation, Formal analysis, Investigation, Methodology, Software, Validation, Visualization, Writing - original draft. **Andrej Jamnik:** Conceptualization, Funding acquisition, Methodology, Resources, Validation, Writing - review & editing. **István Szilágyi:** Funding acquisition, Project administration, Writing - review & editing. **Matija Tomšič:** Conceptualization, Funding acquisition, Project administration, Investigation, Methodology, Supervision, Validation, Writing - original draft, Writing - review & editing.

# References

[1] A. Lajovic, M. Tomšič, A. Jamnik, The complemented system approach: A novel method for calculating the x-ray scattering from computer simulations, J. Chem. Phys. 133(17) (2010) 174123.

[2] J. Cerar, A. Jamnik, I. Pethes, L. Temleitner, L. Pusztai, M. Tomšič, Structural, rheological and dynamic aspects of hydrogen-bonding molecular liquids: Aqueous solutions of hydrotropic tert-butyl alcohol, J. Colloid Interf. Sci. 560 (2020) 730-742.

[3] M. Tomšič, J. Cerar, A. Jamnik, Supramolecular Structure vs. Rheological Properties: 1,4–Butanediol at Room and Elevated Temperatures, J. Colloid Interf. Sci. 557 (2019) 328-335.

[4] J. Cerar, A. Jamnik, M. Tomšič, Supra-molecular structure and rheological aspects of liquid terminal 1,n-diols from ethylene glycol, 1,3-propandiol, 1,4-butanediol to 1,5-pentanediol, J. Mol. Liq. 276 (2019) 307-317.

[5] M. Tomšič, J. Cerar, A. Jamnik, Characterization of the supramolecular assembly in 1,4-butanediol, J. Mol. Liq. 259 (2018) 291-303.

[6] J. Cerar, A. Lajovic, A. Jamnik, M. Tomšič, Performance of various models in structural characterization of n-butanol: Molecular dynamics and X-ray scattering studies, J. Mol. Liq. 229 (2017) 346-357.

[7] M. Tomšič, A. Jamnik, G. Fritz-Popovski, O. Glatter, L. Vlček, Structural properties of pure simple alcohols from ethanol, propanol, butanol, pentanol, to hexanol: Comparing Monte Carlo simulations with experimental SAXS data, J. Phys. Chem. B 111(7) (2007) 1738-1751.

[8] W.L. Hinze, E. Pramauro, A Critical Review of Surfactant-Mediated Phase Separations (Cloud-Point Extractions): Theory and Applications, Crit. Rev. Anal. Chem. 24(2) (1993) 133-177.

[9] A. Patist, S.S. Bhagwat, K.W. Penfield, P. Aikens, D.O. Shah, On the measurement of critical micelle concentrations of pure and technical-grade nonionic surfactants, J. Surfactants Deterg. 3(1) (2000) 53-58.

[10] M.J. Schick, Nonionic surfactants: physical chemistry, CRC Press, 1987.

[11] M. Zulauf, K. Weckstrom, J.B. Hayter, V. Degiorgio, M. Corti, Neutron-Scattering Study of Micelle Structure in Isotropic Aqueous-Solutions of Poly(Oxyethylene) Amphiphiles, J. Phys. Chem. 89(15) (1985) 3411-3417.

[12] G.D. Phillies, R. Hunt, K. Strang, N. Sushkin, Aggregation number and hydrodynamic hydration levels of Brij-35 micelles from optical probe studies, Langmuir 11(9) (1995) 3408-3416.

[13] S. Ghosh, S. Moulik, Interfacial and micellization behaviors of binary and ternary mixtures of amphiphiles (Tween-20, Brij-35, and sodium dodecyl sulfate) in aqueous medium, J. Colloid Interf. Sci. 208(2) (1998) 357-366.

[14] H. Preu, A. Zradba, S. Rast, W. Kunz, E.H. Hardy, M.D. Zeidler, Small angle neutron scattering of D2O-Brij 35 and D2O-alcohol-Brij 35 solutions and their modelling using the Percus-Yevick integral equation, Phys. Chem. Chem. Phys. 1(14) (1999) 3321-3329.

[15] S. Borbély, Aggregate structure in aqueous solutions of Brij-35 nonionic surfactant studied by small-angle neutron scattering, Langmuir 16(13) (2000) 5540-5545.

[16] G. Tóth, Á. Madarász, Structure of BRIJ-35 nonionic surfactant in water: a reverse Monte Carlo study, Langmuir 22(2) (2006) 590-597.

[17] M. Tomšič, M. Bešter-Rogač, A. Jamnik, W. Kunz, D. Touraud, A. Bergmann, O. Glatter, Nonionic surfactant Brij 35 in water and in various simple alcohols: Structural investigations by small-angle X-ray scattering and dynamic light scattering, J. Phys. Chem. B 108(22) (2004) 7021-7032.

[18] M. Tomšič, M. Bešter-Rogač, A. Jamnik, W. Kunz, D. Touraud, A. Bergmann, O. Glatter, Ternary systems of nonionic surfactant Brij 35, water and various simple alcohols: Structural investigations by small-angle X-ray scattering and dynamic light scattering, J. Colloid Interf. Sci. 294(1) (2006) 194-211.

[19] S. Storm, S. Jakobtorweihen, I. Smirnova, Solubilization in Mixed Micelles Studied by Molecular Dynamics Simulations and COSMOmic, J. Phys. Chem. B 118(13) (2014) 3593-3604.

[20] A. Meziani, D. Touraud, A. Zradba, S. Pulvin, I. Pezron, M. Clausse, W. Kunz, Comparison of enzymatic activity and nanostructures in water ethanol Brij 35 and water 1-pentanol Brij 35 systems, J. Phys. Chem. B 101(18) (1997) 3620-3625.

[21] C. Schirmer, Y. Liu, D. Touraud, A. Meziani, S. Pulvin, W. Kunz, Horse liver alcohol dehydrogenase as a probe for nanostructuring effects of alcohols in water/nonionic surfactant system, J. Phys. Chem. B 106(30) (2002) 7414-7421.

[22] A. Lajovic, M. Tomšič, A. Jamnik, Structural Study of Simple Organic Acids by Small-Angle X-Ray Scattering and Monte Carlo Simulations, Acta. Chim. Slov. 59(3) (2012) 520-527.

[23] A. Lajovic, M. Tomšič, G. Fritz-Popovski, L. Vlček, A. Jamnik, Exploring the Structural Properties of Simple Aldehydes: A Monte Carlo and Small-Angle X-Ray Scattering Study, J. Phys. Chem. B 113(28) (2009) 9429-9435.

[24] O. Glatter, Chapter 8 - Numerical Methods, Scattering Methods and their Application in Colloid and Interface Science, Elsevier, 2018, pp. 137-174.

[25] D. Orthaber, A. Bergmann, O. Glatter, SAXS experiments on absolute scale with Kratky systems using water as a secondary standard, J. Appl. Crystallogr. 33 (2000) 218-225.

[26] M.J. Abraham, T. Murtola, R. Schulz, S. Páll, J.C. Smith, B. Hess, E. Lindahl, GROMACS: High performance molecular simulations through multi-level parallelism from laptops to supercomputers, SoftwareX 1-2 (2015) 19-25.

[27] B. Chen, J.J. Potoff, J.I. Siepmann, Monte Carlo Calculations for Alcohols and Their Mixtures with Alkanes. Transferable Potentials for Phase Equilibria. 5. United-Atom Description of Primary, Secondary, and Tertiary Alcohols, The Journal of Physical Chemistry B 105(15) (2001) 3093-3104.

[28] J.M. Stubbs, J.J. Potoff, J.I. Siepmann, Transferable Potentials for Phase Equilibria. 6. United-Atom Description for Ethers, Glycols, Ketones, and Aldehydes, The Journal of Physical Chemistry B 108(45) (2004) 17596-17605.

[29] J.L.F. Abascal, C. Vega, A general purpose model for the condensed phases of water: TIP4P/2005, J. Chem. Phys. 123(23) (2005) 234505.

[30] H.S. Ashbaugh, L. Liu, L.N. Surampudi, Optimization of linear and branched alkane interactions with water to simulate hydrophobic hydration, J. Chem. Phys. 135(5) (2011) 054510.

[31] L. Martinez, R. Andrade, E.G. Birgin, J.M. Martinez, PACKMOL: a package for building initial configurations for molecular dynamics simulations, J Comput Chem 30(13) (2009) 2157-64.

[32] W. Humphrey, A. Dalke, K. Schulten, VMD: Visual molecular dynamics, J Mol Graph Model 14(1) (1996) 33-38.

[33] M.P. Allen, D.J. Tildesley, Computer simulation of liquids, Clarendon Press, Oxford, 2017, pp. 71-110.

[34]     J.G. Kirkwood, F.P. Buff, The Statistical Mechanical Theory of Solutions .1., J. Chem. Phys. 19(6) (1951) 774-777.
[35]     D. Waasmaier, A. Kirfel, New analytical scattering-factor functions for free atoms and ions, Acta Crystallogr. A 51(3) (1995) 416–431.
[36]     P. Debye, Rontgeninterferenzen und Atomgrosse, Phys Z 31 (1930) 419-428.
[37]     N. Dawass, P. Kruger, S.K. Schnell, J.M. Simon, T.J.H. Vlugt, Kirkwood-Buff integrals from molecular simulation, Fluid Phase Equil. 486 (2019) 21-36.
[38]     T.E. Markland, S. Habershon, D.E. Manolopoulos, Quantum diffusion of hydrogen and muonium atoms in liquid water and hexagonal ice, J. Chem. Phys. 128(19) (2008) 194506.
[39]     C. Vega, J.L.F. Abascal, Simulating water with rigid non-polarizable models: a general perspective, Phys. Chem. Chem. Phys. 13(44) (2011) 19663-19688.
[40]     W.S. Price, H. Ide, Y. Arata, Self-Diffusion of Supercooled Water to 238 K Using PGSE NMR Diffusion Measurements, J. Phys. Chem. A 103(4) (1999) 448-450.
[41]     I. Makio, O. Yoko, K. Tadashi, S. Yumi, Y. Kazuhiro, Y. Yoshimi, M. Mitsuo, The Dynamical Structure of Normal Alcohols in Their Liquids as Determined by the Viscosity and Self-Diffusion Measurements, Bull. Chem. Soc. Jpn. 59(12) (1986) 3771-3774.
[42]     E.J.W. Wensink, A.C. Hoffmann, P.J. van Maaren, D. van der Spoel, Dynamic properties of water/alcohol mixtures studied by computer simulation, J. Chem. Phys. 119(14) (2003) 7308-7317.
[43]     S.K. Vahvaselka, R. Serimaa, M. Torkkeli, Determination of liquid structures of the primary alcohols methanol, ethanol, l-propanol, 1-butanol and 1-octanol by x-ray scattering, J. Appl. Crystallogr. 28 (1995) 189-195.
[44]     W. Heller, Small-angle neutron scattering and contrast variation: a powerful combination for studying biological structures, Acta Crystallogr., Sect. D: Biol. Crystallogr. 66(11) (2010) 1213-1217.
[45]     C. Haber, S.A. Ruiz, D. Wirtz, Shape anisotropy of a single random-walk polymer, Proc. Natl. Acad. Sci. U. S. A. 97(20) (2000) 10792-10795.
[46]     H. Lee, R.M. Venable, A.D. MacKerell Jr, R.W. Pastor, Molecular dynamics studies of polyethylene oxide and polyethylene glycol: hydrodynamic radius and shape anisotropy, Biophys. J. 95(4) (2008) 1590-1599.
[47]     R.C. Pasquali, M.P. Taurozzi, C. Bregni, Some considerations about the hydrophilic–lipophilic balance system, Int. J. Pharm. 356(1) (2008) 44-51.
[48]     W.S. Price, H. Ide, Y. Arata, Solution dynamics in aqueous monohydric alcohol systems, J. Phys. Chem. A 107(24) (2003) 4784-4789.
[49]     F.M. Samigullin, Study of translational self-diffusion of molecules in liquids, Journal of Structural Chemistry 14(4) (1974) 569-574.
[50]     R.E. Rathbun, A.L. Babb, Self-Diffusion in Liquids. 3. Temperature Dependence in Pure Liquids, J. Phys. Chem. 65(6) (1961) 1072-1074.
[51]     D.W. Mccall, D.C. Douglass, Self-Diffusion in the Primary Alcohols, J. Chem. Phys. 32(6) (1960) 1876-1877.
[52]     K.R. Harris, P.J. Newitt, Z.J. Derlacki, Alcohol tracer diffusion, density, NMR and FTIR studies of aqueous ethanol and 2,2,2-trifluoroethanol solutions at 25 degrees C, J. Chem. Soc. Faraday Trans. 94(14) (1998) 1963-1970.
[53]     A. Klinov, I. Anashkin, Diffusion in Binary Aqueous Solutions of Alcohols by Molecular Simulation, Processes 7(12) (2019) 947.
[54]     J.T. Su, P.B. Duncan, A. Momaya, A. Jutila, D. Needham, The effect of hydrogen bonding on the diffusion of water in n-alkanes and n-alcohols measured with a novel single microdroplet method, J. Chem. Phys. 132(4) (2010) 044506.

[55] M.A. Lusis, G.A. Ratcliff, Diffusion of Inert and Hydrogen-Bonding Solutes in Aliphatic Alcohols, Aiche Journal 17(6) (1971) 1492-1496.
[56] Z.L. Wang, D. Afanasenkau, M.J. Dong, D.N. Huang, S. Wiegand, Molar mass and temperature dependence of the thermodiffusion of polyethylene oxide in water/ethanol mixtures, J. Chem. Phys. 141(6) (2014) 064904.
[57] R. Kita, S. Wiegand, J. Luettmer-Strathmann, Sign change of the Soret coefficient of poly(ethylene oxide) in water/ethanol mixtures observed by thermal diffusion forced Rayleigh scattering, J. Chem. Phys. 121(8) (2004) 3874-3885.
[58] B.J. de Gans, R. Kita, S. Wiegand, J. Luettmer-Strathmann, Unusual thermal diffusion in polymer solutions, Physical Review Letters 91(24) (2003) 245501.
[59] K. Shimada, H. Kato, T. Saito, S. Matsuyama, S. Kinugasa, Precise measurement of the self-diffusion coefficient for poly(ethylene glycol) in aqueous solution using uniform oligomers, J. Chem. Phys. 122(24) (2005) 244914.
[60] J.S. Hub, Interpreting solution X-ray scattering data using molecular simulations, Curr Opin Struc Biol 49 (2018) 18-26.
[61] M. Tomšič, G. Fritz-Popovski, L. Vlček, A. Jamnik, Calculating small-angle x-ray scattering intensities from Monte Carlo results: Exploring different approaches on the example of primary alcohols, Acta. Chim. Slov. 54(3) (2007) 484–491.
[62] D. Frenkel, V.R. J., C.G. de Kruif, A. Vrij, Structure factors of polydisperse systems of hard spheres: A comparison of Monte Carlo simulations and Percus-Yevick theory, J. Chem. Phys. 84(8) (1986) 4625-4630.
[63] K. Schmidt-Rohr, Simulation of small-angle scattering curves by numerical Fourier transformation, J. Appl. Crystallogr. 40 (2007) 16-25.
[64] P. Debye, A.M. Bueche, Scattering by an Inhomogeneous Solid, J. Appl. Phys. 20(6) (1949) 518-525.
[65] J.S. Pedersen, Analysis of small-angle scattering data from colloids and polymer solutions: modeling and least-squares fitting, Adv. Col. Interf. Sci. 70 (1997) 171-210.
[66] D. Svergun, C. Barberato, M.H.J. Koch, CRYSOL - A program to evaluate x-ray solution scattering of biological macromolecules from atomic coordinates, J. Appl. Crystallogr. 28 (1995) 768-773.
[67] B.C. McAlister, B.P. Gradely, The Use of Monte Carlo Simulations to Calculate Small-Angle Scattering Patterns, Macromol. Symp. 190 (2002) 117-129.
[68] R.L. McGreevy, L. Pusztai, Reverse Monte Carlo Simulation: A New Technique for the Determination of Disordered Structures, Mol. Simul. 1(6) (1988) 359-367.
[69] S. Brisard, P. Levitz, Small-angle scattering of dense, polydisperse granular porous media: Computation free of size effects, Phys. Rev. E 87(1) (2013) 013305.
[70] K. Hinsen, E. Pellegrini, S. Stachura, G.R. Kneller, nMoldyn 3: Using task farming for a parallel spectroscopy-oriented analysis of molecular dynamics simulations, J Comput Chem 33(25) (2012) 2043-2048.
[71] S. Schottl, T. Lopian, S. Prevost, D. Touraud, I. Grillo, O. Diat, T. Zemb, D. Horinek, Combined molecular dynamics (MD) and small angle scattering (SAS) analysis of organization on a nanometer-scale in ternary solvent solutions containing a hydrotrope, J. Colloid Interf. Sci. 540 (2019) 623-633.
[72] B. Chen, J.J. Potoff, J.I. Siepmann, Monte Carlo calculations for alcohols and their mixtures with alkanes. Transferable potentials for phase equilibria. 5. United-atom description of primary, secondary, and tertiary alcohols, J. Phys. Chem. B 105(15) (2001) 3093-3104.
[73] J.M. Stubbs, J.J. Potoff, J.I. Siepmann, Transferable potentials for phase equilibria. 6. United-atom description for ethers, glycols, ketones, and aldehydes, J. Phys. Chem. B 108(45) (2004) 17596-17605.

[74] S. Henkel, M.C. Misuraca, P. Troselj, J. Davidson, C.A. Hunter, Polarisation effects on the solvation properties of alcohols, Chem Sci 9(1) (2018) 88-99.
[75] V. Makarov, B.M. Pettitt, M. Feig, Solvation and Hydration of Proteins and Nucleic Acids:  A Theoretical View of Simulation and Experiment, Acc. Chem. Res. 35(6) (2002) 376-384.
[76] P. Ren, J. Chun, D.G. Thomas, M.J. Schnieders, M. Marucho, J. Zhang, N.A. Baker, Biomolecular electrostatics and solvation: a computational perspective, Q. Rev. Biophys. 45(4) (2012) 427-491.
[77] H.M. Berman, Hydration of DNA: take 2, Curr Opin Struc Biol 4(3) (1994) 345-350.
[78] H.M. Berman, Hydration of DNA, Curr Opin Struc Biol 1(3) (1991) 423-427.
[79] J.A. Long, B.M. Rankin, D. Ben-Amotz, Micelle Structure and Hydrophobic Hydration, J. Am. Chem. Soc. 137(33) (2015) 10809-10815.
[80] K. Streletzky, G.D. Phillies, Temperature dependence of Triton X-100 micelle size and hydration, Langmuir 11(1) (1995) 42-47.
[81] E.M. Corbeil, N.E. Levinger, Dynamics of Polar Solvation in Quaternary Microemulsions, Langmuir 19(18) (2003) 7264-7270.
[82] D. Mandal, A. Datta, S.K. Pal, K. Bhattacharyya, Solvation Dynamics of 4-Aminophthalimide in Water-in-Oil Microemulsion of Triton X-100 in Mixed Solvents, J. Phys. Chem. B 102(45) (1998) 9070-9073.
[83] M. Allen, D. Linder, Ethylene oxide oligomer distribution in nonionic surfactants via high performance liquid chromatography (HPLC), J. Am. Oil Chem. Soc. 58(10) (1981) 950-957.
[84] K. Nakamura, Y. Morikawa, I. Matsumoto, Rapid analysis of ionic and nonionic surfactant homologs by high performance liquid chromatography, J. Am. Oil Chem. Soc. 58(1) (1981) 72-77.
[85] C. Crescenzi, A. Di Corcia, R. Samperi, A. Marcomini, Determination of Nonionic Polyethoxylate Surfactants in Environmental Waters by Liquid Chromatography/Electrospray Mass Spectrometry, Anal. Chem. 67(11) (1995) 1797-1804.
[86] J. Chen, S.K. Spear, J.G. Huddleston, R.D. Rogers, Polyethylene glycol and solutions of polyethylene glycol as green reaction media, Green Chemistry 7(2) (2005) 64-82.
[87] B.R. Venugopal, C. Shivakumara, M. Rajamathi, Effect of various factors influencing the delamination behavior of surfactant intercalated layered double hydroxides, J. Colloid Interf. Sci. 294(1) (2006) 234-239.
[88] J.T. Rajamathi, N. Ravishankar, M. Rajamathi, Delamination-restacking behaviour of surfactant intercalated layered hydroxy double salts, M3Zn2(OH)(8)(surf)(2)center dot 2H(2)O [M = Ni, Co and surf = dodecyl sulphate (DS), dodecyl benzene sulphonate (DBS)], Solid State Sci 7(2) (2005) 195-199.

Jure Cerar,[a] Andrej Jamnik,[a] István Szilágyi,[b] and Matija Tomšič.[a,*]

[a] *Faculty of Chemistry and Chemical Technology, University of Ljubljana, Večna pot 113, SI-1000 Ljubljana, Slovenia.*
[b] *MTA-SZTE Lendület Biocolloids Research Group, Interdisciplinary Excellence Center, Department of Physical Chemistry and Materials Science, University of Szeged, H-6720 Szeged, Hungary.*
*Correspondence e-mail: Matija.Tomšič@fkkt.uni-lj.si; tel.: +386 1 479 8515 (M. Tomšič)*

## APPENDIX A - Experimental and Methods

**Small- and Wide-Angle X-Ray Scattering**

The SWAXS measurements were performed in an in-lab-modified old-Kratky-type camera (Anton Paar KG, Graz, Austria) equipped with a conventional x-ray generator (GE Inspection Technologies, SEIFERT ISO-DEBYEFLEX 3003) with a Cu-anode operating at 40 kV and 50 mA (Cu-$K_\alpha$ line with the wavelength $\lambda$ = 1.54 Å). Focusing multilayer optics (Göbel mirror) were used to focus and monochromatize the primary x-ray beam. The SWAXS camera was equipped with a block-collimation unit to provide a well-defined, line-collimated primary beam. Therefore, the resulting SWAXS data were experimentally smeared. The samples were placed in a standard quartz capillary (outer diameter of 1 mm and wall thickness of 10 μm) and thermostated at 25 °C using a Peltier element. SWAXS measurements were recorded on a 2D-imaging plate, which was exposed to the scattered x-rays for 60 min, and read-off afterwards utilizing a Fuji BAS 1800II imaging-plate reader with a spatial resolution of 50×50 μm$^2$/px. The SWAXS data were obtained in the range of the scattering vector, $q$, values from 0.1 to 25 nm$^{-1}$, where $q = 4\pi/\lambda \cdot \sin(\theta/2)$, with $\theta$ being the scattering angle. These data were then corrected for the sample x-ray absorption and the background scattering and further transformed to the absolute scale using water as a secondary standard [1]. The resulting SWAXS data were still experimentally smeared due to the finite dimensions of the primary beam [2].

**Molecular Dynamics Simulations**

All the simulations were carried out using the GROMACS 2018.6 software package [3]. The Brij-35 and alcohol molecules were described according to the TraPPE-UA force field model [4, 5] and the water molecules by the TIP4P/2005 force-field model for water [6]. Additionally, in our simulations we applied the alkane-site/water-oxygen interaction correction proposed by Ashbaugh *et al* [7]. The MD simulations were conducted in isothermal–isobaric ensemble ($NPT$) at 300 K and 100 kPa. We used the Nose-Hoover temperature coupling and Parrinello-Rahman pressure coupling with a relaxation time of 3 ps. The equations of motion were solved using the Verlet leapfrog algorithm. The short-range interactions were calculated using the Verlet neighbor scheme with a single cut-off distance of 1.4 nm and long-range dispersion corrections for the energy and pressure. The long-range Coulombic interactions were handled by the smooth particle-mesh Ewald (PME). The initial configurations were constructed using Packmol software (version 18.169) [8] by placing the molecules in a simulation box semi-randomly (accounting for the excluded volume). The number of molecules ($N_{Brij35}$ ~ 1200, $N_{water}$ ~ 32000, $N_{Alc}$ ~ 580000–260000) was selected to match the selected mass fraction of the solutes and in accordance with the simulation box's side lengths of approximately 40 nm.

The simulations were performed with a time step of 2 fs for 7×10$^6$ steps, representing a time scale of 14 ns. The first 4 ns of the simulation were discarded, leaving 10 ns for the statistical and structural analyses of the model system. A configuration of the simulation box was saved every 10 ps, resulting in 1000 saved configurations for the subsequent statistical analysis and the SWAXS intensity calculation. The presented snapshots of the simulation-box configurations and molecular conformations were rendered using VMD software [9].

**Calculation of the X-Ray Scattering Intensities**

The theoretical SWAXS intensities were calculated from 1000 independent configurations of the simulation box, which were obtained from the MD simulations, using the complemented-system approach that is explained in detail elsewhere [10-12]. The atomic form factors used in these calculations were calculated using the analytical expressions [13], while the form factors for the pseudo-atoms were obtained utilizing the Debye equation [10, 12].

The resulting calculated SWAXS intensities were further numerically smeared utilizing the experimental primary beam 'width' and 'length' profiles and were further directly compared to the experimental SWAXS data on an absolute scale.

Slight oscillations can be observed in the very low-*q* region (below ~ 1 nm$^{-1}$) of some calculated SWAXS curves (see Fig. 1b and 1c), which arise from the fact that the inter-molecular correlations do not fully decay over the distance of the simulation box's half length. However, we need to make a compromise between increasing the size of the simulation box's side length and the computing time necessary for the MD simulations and calculations of the scattering intensities. Correspondingly, our simulation box sizes were around 40 nm.

**Calculation of the Self-Diffusion Coefficients**

The self-diffusion coefficients, $D_i$, were calculated from the MD-simulation results on the mean-square displacement of the particles using the following expression [14]:

$$\lim_{t\to\infty}\left\langle \left|\vec{r}_i(t)-\vec{r}_i(0)\right|^2 \right\rangle = b_o + 6D_i \cdot t, \quad (S1)$$

where $\vec{r}_i(0)$ and $\vec{r}_i(t)$ are the initial position of the $i$-th particle in the system and its position at time $t$, respectively, and $b_0$ the regression parameter. For this purpose the particle positions were sampled every 0.2 ps during the 2 ns of the time scale of the system.

**Calculation of the Kirkwood-Buff Integrals**

The Kirkwood-Buff integrals (KBIs), $G_{ij}$, were calculated from the radial-distribution functions (RDFs) obtained as a result of the MD simulations. Assuming spherical symmetry, the KBIs can be calculated as follows [15]:

$$G_{ij} = 4\pi \int_0^\infty \left(g_{ij}(r)-1\right) r^2 \, dr, \quad (S2)$$

where $g_{ij}(r)$ is the radial distribution function between the particles of $i$-th and $j$-th species. Since this equation is only valid for infinite systems, we have accounted for finite distances by applying the correction term and calculating [16]:

$$G_{ij} = 4\pi \int_0^\infty \left(g_{ij}(r)-1\right) r^2 \left(1 - \frac{3x}{2} + \frac{x^3}{2}\right) dr, \quad (S3)$$

where $x$ is a dimensionless distance, $x = r/L$, where $L$ is half the side-length of the simulation box.

## APPENDIX B - Results and Discussion

### Experimental and Calculated SWAXS Data

In figures presenting the SWAXS intensity results the logarithmic scale is used on the ordinate axis as usual when presenting scattering data to facilitate the visibility of the scattering details across many orders of magnitude. The calculated SWAXS intensities in Fig. 1 (thin lines) are numerically smeared and as such are directly comparable to the experimental SWAXS data (thick lines). The uncertainties of both the experimental and calculated scattering-intensity data are well below the line thickness.

### Inter-Molecular Structure

In this study we are especially interested in the molecular organization of the nonionic surfactant in alcohol and alcohol-rich media and the distribution of the water molecules (hydration) in these systems. The structural situations found in the modelled 5 *wt.* % Brij 35/alcohol and 5 *wt.* % Brij 35/2 *wt.* % water/alcohol systems is presented in Fig. S1. The presence of solvated and hydrated Brij 35 monomers is confirmed in these systems – even though at first sight it would be possible to claim that a slight aggregation of the surfactant molecules can be observed, a more careful observation reveals that the individual surfactant molecules are, in the great majority of cases, surrounded by at least one 'layer' of alcohol molecules. Furthermore, it turns out that the system is highly dynamic. Therefore, even if occasionally (statistically speaking) an aggregate is formed, it is not long-lived.

The detailed partial scattering contributions to the theoretical SWAXS intensity of the Brij 35/water/HexOH system are presented in Fig. S2 together with the cross-term contribution. The details of how we obtained these contributions according to the complemented-system approach are described elsewhere [17]. At this point we need to comment on the fine oscillations that can be observed on some curves at low values of the scattering vector in Fig. 3 and Fig. S2 – they arise from the somewhat worse statistics due to the small number of surfactant molecules in the simulation box as the box size is limited by the available computing power. Nevertheless, these fine oscillations obviously do not prevent us from drawing reasonable conclusions based on these data, which are discussed in the paper.

We continue the discussion with the interesting structural features connected to the hydration of the Brij35 hydrophilic heads, which can be observed from the radial distribution functions shown in Fig. 4b and 4c. When considering the anisotropic Brij 35 molecule, we could only calculate its centre-of-mass (COM) and took it as a reference point for calculating the corresponding PDFs. Moreover, the Brij 35 molecule cannot be considered a "hard" molecule (i.e., a 'hard sphere') with respect to water molecules, since water molecules can easily penetrate and hydrate its hydrophilic poly(oxyethylene) part, even in close proximity to its COM. Therefore, the COM-water PDFs look rather atypical with their central maximum around *r*-value zero. We must also point out that due to the relatively large surfactant hydrophilic head the Brij 35's COM is actually located very close to the COM of the Brij 35 hydrophilic head. The COM–AlcOH and COM–AlcCH$_x$ distribution functions provide a deeper insight into the solvation of surfactant molecules in alcohols. We can see that all the functions show shallow, broad minima at short distances with values below one. This means a rather low probability of the presence of the neighbouring alcohol molecule at these distances, which is a direct consequence of the excluded volume of the hydrophilic head of the surfactant molecule in the hydrophobic environment represented by the hydrocarbon tails of the alcohol molecules. Interestingly, with an increase in the hydrophobicity of the solvent (alcohol), which causes an increase in the local segregation of the hydrophilic and hydrophobic regions, we can observe a double minimum in the COM–AlcOH distribution functions in this range, *i.e.*, a steady increase in the functional course up to ~0.7 nm and a further decrease with a shallow second minimum at ~1.3 nm. This feature can be explained by the orientation of the alcohol molecules, with the hydroxyl groups towards the surfactant hydrophilic head, as illustrated in the scheme in Fig. 4. Correspondingly, there is a lower probability of finding another alcohol –OH group in the intermediate range of distances between 0.7 nm and 1.8 nm, as this range corresponds to the hydrophobic regions occupied by the alcohol alkyl tails. Such an explanation is justified for the HexOH system, where these minima are the most clearly expressed in Fig. 4b. We can also observe a slight shift of the maximum around 2 nm towards somewhat larger distances with the presence of water in the system, which is probably a consequence of the bulkier, hydrated Brij 35 hydrophilic head. Interestingly, the separation of the corresponding maxima in the $g(r)$ function is about 1.4 nm, which conforms to the distance between the neighbouring –OH skeletons in pure HexOH (see Figs. 3d and 3e in ref. [18] and ref. [17]). Similarly, the position of the intermediate maximum in the COM–AlcOH distribution function of the HexOH system in Fig. 4b provides us with a rough estimate of the effective radius of the Brij 35 hydrophilic head, which is around 0.7 nm.

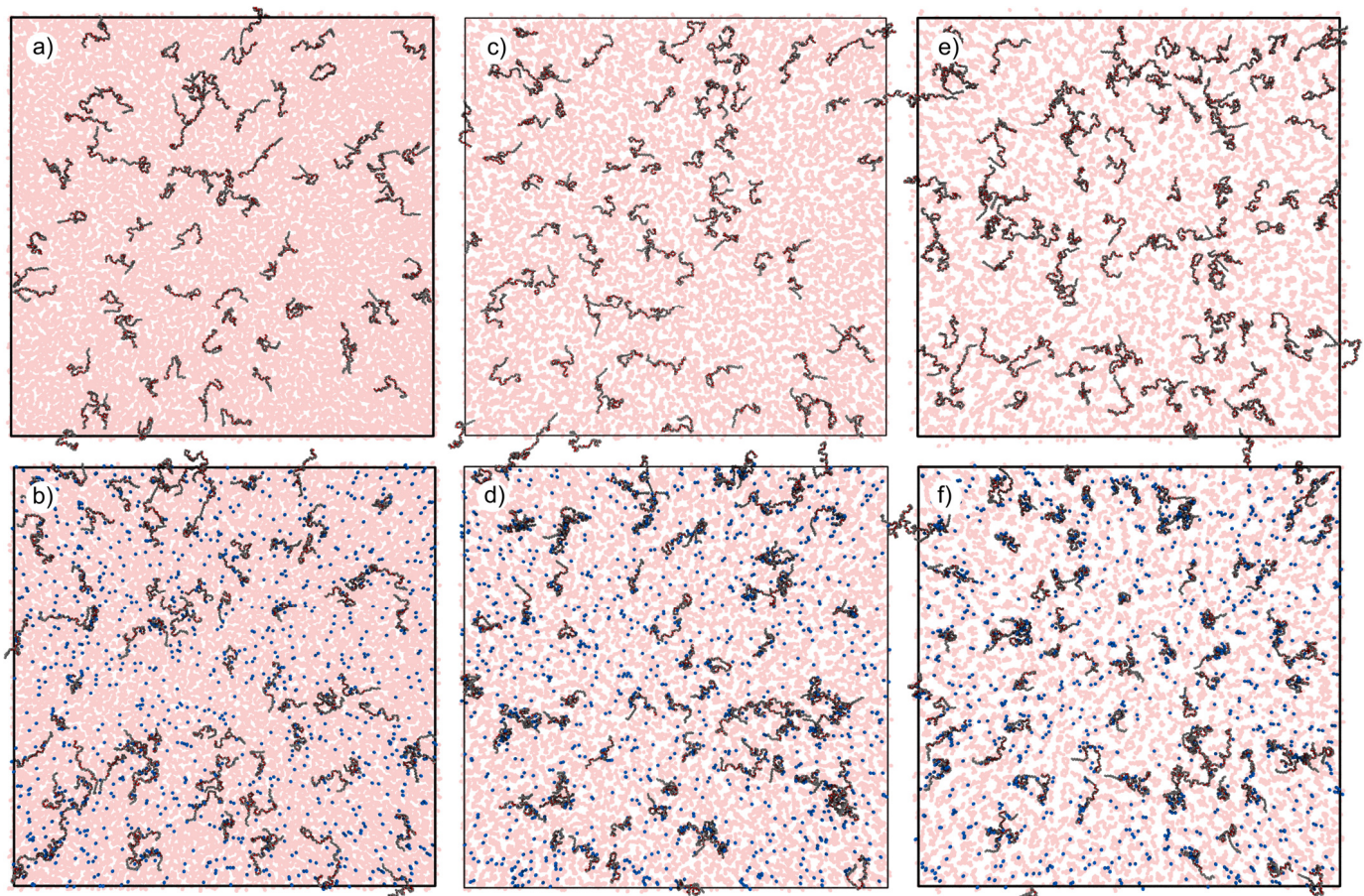

**Fig. S1.** F Visualization of 1-nm-thick 40 nm × 40 nm slice of the MD-simulation box for the studied (a) Brij 35 / EtOH, (b) Brij 35 / water / EtOH, (c) Brij 35/BuOH, (d) Brij 35/water/BuOH, (e) Brij 35 / HexOH, and (f) Brij 35/water/HexOH systems, where the alcohol molecules are omitted for the sake of clarity – the pale red areas indicate the hydrophilic regions formed by the sequentially hydrogen-bonded alcohol hydroxyl groups; the water molecules are depicted in blue.

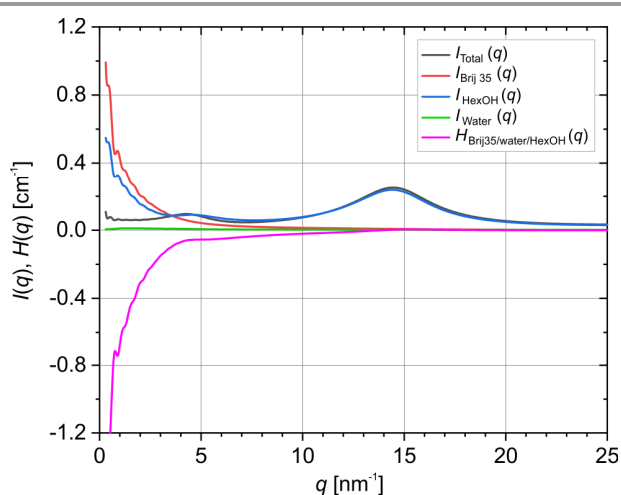

**Fig. S2.** Partial scattering contributions (not smeared) to the theoretical SWAXS intensitiy of the Brij 35/water/HexOH system: Brij 35 contribution (red), HexOH contribution (blue), water contribution (green), and the scattering cross term contribution (magenta). The total scattering intensitiy of the Brij 35/water/HexOH system is also shown (black).

**Values of the Kirkwood-Buff Integrals**

The values of the calculated Kirkwood-Buff Integrals for different chosen pair-distribution functions are gathered in Table S1. The KBI, $G_{ij}$, is a quantity that can be used as a measure of the affinity between molecules of the *i*-th and *j*-th species in solution. Its positive value can generally be interpreted as the excess of component *j* around component *i* and, on the in contrary, its negative value as the depletion of component *j* around component *i*. Some interesting trends in the KBI values upon addition of small concentration of water to the Brij 35/alcohol binary system or upon changing the length of the hydrocarbon tail of the alcohol molecule can be observed in Table S1. E.g., the values of $G_{COM-water}$ decrease to strongly negative values with increasing length of the hydrocarbon tail of the alcohol (increasing hydrophobic nature of the alcohol), which we can interpret as a direct consequence of increasing penetration of water into the hydrophilic Brij 35 head (hydration) and an associated depletion of water molecules at greater distances from COM. It is also interesting to note that when water is added to the Brij 35/alcohol system, the $G_{COM-AlcOH}$ value actually increases slightly in the case of (hydrophilic) ethanol, but decreases significantly in the case of more hydrophobic butanol and hexanol. Obviously, in hydrophobic alcohols the alcohol –OH group shows a stronger depletion around the Brij 35 COM, probably the consequence of the more hydrated hydrophilic Brij 35 head and the consequent slightly stronger segregation of hydrophilic and hydrophobic regions in these systems. Virtually the same trend with very similar justification, can also be observed for the trend in $G_{COM-AlcCH_x}$ values, corresponding to the depletion of hydrophobic hydrocarbon tails around the COM of Brij 35. However, the trend in $G_{COM-COM}$ values arises from the fact that Brij 35 molecules adopt more compact conformations when hydrated in more hydrophobic alcohols (see also Fig. S1 and Fig. 6).

**Intra-Molecular Structure**

In order to obtain effective general information about the average shape of the modelled surfactant molecule presented in Fig. 5, all the molecules in the simulation box were put into the same reference frame and analysed geometrically. In our case this involved finding the COM and the vector $r$ along the longest intra-molecular distance from the hydrophilic head to the hydrophobic tail of the surfactant molecule, and further rotating each molecule in such a way that the vector $r$ was in parallel with the *x*-axis, as depicted schematically in the upper part of Fig. 5.

To calculate the distribution of the water molecules around the COM presented in Fig. 5 (blue curve), which clearly shows intensive hydration of the hydrophilic surfactant head we only considered the water molecules that are hydrogen-bonded to the Brij 35 molecules, *i.e.*, the water molecules within the cut-off distance of 1.6 nm from the COM (this cut-off distance was obtained from the first minimum in the COM-water $g(r)$ function in Fig. 4a).

**Table S1.** Values of selected Kirkwood-Buff integrals for the binary and ternary systems studied. All values are in units of nm³.

| System | $G_{COM-water}$ | System | $G_{COM-water}$ |
|---|---|---|---|
| EtOH/Brij 35 | - | EtOH/Brij 35/water | 2.920 |
| BuOH/Brij 35 | - | BuOH/Brij 35/water | -0.591 |
| HexOH/Brij 35 | - | HexOH/Brij 35/water | -4.304 |
| System | $G_{COM-AlcOH}$ | System | $G_{COM-AlcOH}$ |
| EtOH/Brij 35 | -2.703 | EtOH/Brij 35/water | -2.358 |
| BuOH/Brij 35 | -1.716 | BuOH/Brij 35/water | -2.398 |
| HexOH/Brij 35 | -0.425 | HexOH/Brij 35/water | -2.72 |
| System | $G_{COM-AlcCH_x}$ | System | $G_{COM-AlcCH_x}$ |
| EtOH/Brij 35 | -2.709 | EtOH/Brij 35/water | -2.365 |
| BuOH/Brij 35 | -1.737 | BuOH/Brij 35/water | -2.420 |
| HexOH/Brij 35 | -0.460 | HexOH/Brij 35/water | -2.774 |
| System | $G_{COM-COM}$ | System | $G_{COM-COM}$ |
| EtOH/Brij 35 | 32.29 | EtOH/Brij 35/water | 29.03 |
| BuOH/Brij 35 | 15.45 | BuOH/Brij 35/water | 30.80 |
| HexOH/Brij 35 | -31.58 | HexOH/Brij 35/water | 37.71 |

The results in Fig. 6a also reveal the effects of the presence of water on the Brij 35's effective molecular conformations. The distribution *x-y* cross-section profile narrows slightly in the case of EtOH, stays similar in the case of BuOH, and widens slightly in case of the HexOH ternary system with water. These changes are solely due to the hydration of the hydrophilic part of the surfactant molecule. The hydrating water molecules hydrogen bond to the hydrophilic chain in the Brij 35's head and act as the intra-molecular bridges, as is clearly depicted in Fig. 2c. Such bridging can also lead to the compaction of the surfactant hydrophilic head, which can explain the observed narrowing of the distribution profile in the EtOH system – an increasing hydration of the surfactant hydrophobic head with an increasing hydrophobicity of the solvent would act in the opposite direction (swelling). We would of course expect that such bridging should also influence the dynamic properties of these molecules, which will be discussed in the following subsection.

## References

[1] D. Orthaber, A. Bergmann, O. Glatter, SAXS experiments on absolute scale with Kratky systems using water as a secondary standard, J. Appl. Crystallogr. 33 (2000) 218-225.

[2] O. Glatter, Chapter 8 - Numerical Methods, Scattering Methods and their Application in Colloid and Interface Science, Elsevier, 2018, pp. 137-174.

[3] M.J. Abraham, T. Murtola, R. Schulz, S. Páll, J.C. Smith, B. Hess, E. Lindahl, GROMACS: High performance molecular simulations through multi-level parallelism from laptops to supercomputers, SoftwareX 1-2 (2015) 19-25.

[4] B. Chen, J.J. Potoff, J.I. Siepmann, Monte Carlo Calculations for Alcohols and Their Mixtures with Alkanes. Transferable Potentials for Phase Equilibria. 5. United-Atom Description of Primary, Secondary, and Tertiary Alcohols, The Journal of Physical Chemistry B 105(15) (2001) 3093-3104.

[5] J.M. Stubbs, J.J. Potoff, J.I. Siepmann, Transferable Potentials for Phase Equilibria. 6. United-Atom Description for Ethers, Glycols, Ketones, and Aldehydes, The Journal of Physical Chemistry B 108(45) (2004) 17596-17605.

[6] J.L.F. Abascal, C. Vega, A general purpose model for the condensed phases of water: TIP4P/2005, J. Chem. Phys. 123(23) (2005) 234505.

[7] H.S. Ashbaugh, L. Liu, L.N. Surampudi, Optimization of linear and branched alkane interactions with water to simulate hydrophobic hydration, J. Chem. Phys. 135(5) (2011) 054510.

[8] L. Martinez, R. Andrade, E.G. Birgin, J.M. Martinez, PACKMOL: a package for building initial configurations for molecular dynamics simulations, J Comput Chem 30(13) (2009) 2157-64.

[9] W. Humphrey, A. Dalke, K. Schulten, VMD: Visual molecular dynamics, J Mol Graph Model 14(1) (1996) 33-38.

[10] A. Lajovic, M. Tomšič, A. Jamnik, The complemented system approach: A novel method for calculating the x-ray scattering from computer simulations, J. Chem. Phys. 133(17) (2010) 174123.

[11] J. Cerar, A. Jamnik, I. Pethes, L. Temleitner, L. Pusztai, M. Tomšič, Structural, rheological and dynamic aspects of hydrogen-bonding molecular liquids: Aqueous solutions of hydrotropic tert-butyl alcohol, J. Colloid Interf. Sci. 560 (2020) 730-742.

[12] P. Debye, Rontgeninterferenzen und Atomgrosse, Phys Z 31 (1930) 419-428.

[13] D. Waasmaier, A. Kirfel, New analytical scattering-factor functions for free atoms and ions, Acta Crystallogr. A 51(3) (1995) 416–431.

[14] M.P. Allen, D.J. Tildesley, Computer simulation of liquids, Clarendon Press, Oxford, 2017, pp. 71-110.

[15] J.G. Kirkwood, F.P. Buff, The Statistical Mechanical Theory of Solutions .1., J. Chem. Phys. 19(6) (1951) 774-777.

[16] N. Dawass, P. Kruger, S.K. Schnell, J.M. Simon, T.J.H. Vlugt, Kirkwood-Buff integrals from molecular simulation, Fluid Phase Equil. 486 (2019) 21-36.

[17] M. Tomšič, A. Jamnik, G. Fritz-Popovski, O. Glatter, L. Vlček, Structural properties of pure simple alcohols from ethanol, propanol, butanol, pentanol, to hexanol: Comparing Monte Carlo simulations with experimental SAXS data, J. Phys. Chem. B 111(7) (2007) 1738-1751.

[18] J. Cerar, A. Jamnik, M. Tomsic, Supra-molecular structure and rheological aspects of liquid terminal 1, n-diols from ethylene glycol, 1,3-propandiol, 1,4-butanediol to 1,5-pentanediol, J. Mol. Liq. 276 (2019) 307-317.